%% file: paper.tex
\documentclass[sigconf]{acmart}
\renewcommand\footnotetextcopyrightpermission[1]{}
\def\BibTeX{{\rm B\kern-.05em{\sc i\kern-.025em b}\kern-.08emT\kern-.1667em\lower.7ex\hbox{E}\kern-.125emX}}

\setcopyright{none}

\input{text/preamble}
\input{text/macros}

\input{text/acronyms}

\begin{document}

\title{Object-Capability as a Means of Permission and Authority in Software Systems}

\author{J\"{o}rn Koepe}
\email{jkoepe@mail.uni-paderborn.de}
\affiliation{
  \institution{Paderborn University}
  \country{Germany}
}

\input{text/title_page}

\clearpage
\pagenumbering{arabic}
\input{text/abstract}

%
%



\maketitle

\sloppy

\input{text/body}

\begin{acks}
This research was conducted under the supervision of Ben Hermann as part of the Secure Systems Engineering seminar at Paderborn University, organized by Eric Bodden and Lisa Nguyen Quang Do. It was partially funded by the Heinz Nixdorf Foundation.
\end{acks}

\bibliographystyle{ACM-Reference-Format}
\bibliography{paper}

\end{document}

%% file: text/preamble.tex
\usepackage{xspace}
\usepackage{listings}
\usepackage{tikz}
\usepackage[noend]{algpseudocode}
\usepackage{tabularx,booktabs}
\usepackage[T1]{fontenc}
\usepackage[english]{babel}
\usepackage{pdfpages}
\usepackage{adjustbox}
\usepackage[bookmarks=false]{hyperref}

\usepackage[nolist]{acronym}

%% file: text/macros.tex
\let\oldReturn\Return
\renewcommand{\Return}{\State\oldReturn}

\definecolor{darkred}{rgb}{0.75,0,0}
\definecolor{darkblue}{rgb}{0,0,0.75}
\definecolor{darkgreen}{rgb}{0,0.75,0}
\definecolor{ncs}{rgb}{0.0, 0.53, 0.74}

\newcommand{\hide}[1]{}

\definecolor{pblue}{rgb}{0.13,0.13,1}
\definecolor{pgreen}{rgb}{0,0.5,0}
\definecolor{pred}{rgb}{0.9,0,0}
\definecolor{pgrey}{rgb}{0.46,0.45,0.48}
\algnewcommand\algorithmicswitch{\textbf{switch}}
\algnewcommand\algorithmiccase{\textbf{case}}
\algnewcommand\algorithmicassert{\texttt{assert}}
\algnewcommand\Assert[1]{\State \algorithmicassert(#1)}%
\algdef{SE}[SWITCH]{Switch}{EndSwitch}[1]{\algorithmicswitch\ #1\ \algorithmicdo}{\algorithmicend\ \algorithmicswitch}%
\algdef{SE}[CASE]{Case}{EndCase}[1]{\algorithmiccase\ #1}{\algorithmicend\ \algorithmiccase}%
\algtext*{EndSwitch}%
\algtext*{EndCase}%

\usetikzlibrary{fit,calc,trees,positioning,arrows,chains,shapes.geometric,decorations.pathreplacing,decorations.pathmorphing,shapes,matrix,shapes.symbols, automata}
\tikzset{
>=stealth',
  punktchain/.style={
    rectangle,
    rounded corners,
    draw=black, very thick,
    text width=5em,
    minimum height=3em,
    text centered,
    on chain},
  line/.style={draw, thick, <-},
  element/.style={
    tape,
    top color=white,
    bottom color=blue!50!black!60!,
    minimum width=5em,
    draw=blue!40!black!90, very thick,
    text width=5em,
    minimum height=3.5em,
    text centered,
    on chain},
  every join/.style={->, thick,shorten >=1pt},
  tuborg/.style={decorate},
  tubnode/.style={midway, right=2pt},
}

%% file: text/acronyms.tex
\newacro{EW}{Eventual World}
\newacro{BW}{Behaviour World}
\newacro{MW}{Maximal World}
\newacro{CP}{Current Permission}
\newacro{EP}{Eventual Permission}
\newacro{BP}{Behaviour Permission}
\newacro{MP}{Maximal Permission}

\newacro{CA}{Current Authority}
\newacro{EA}{Eventual Authority}
\newacro{BA}{Behaviour Authority}
\newacro{MA}{Maximal Authority}

%% file: text/title_page.tex
\thispagestyle{empty}

\begin{center}	

	\colorbox{ncs}{
		\begin{minipage}{17cm}
			\begin{minipage}{.68\textwidth}
  			{\color{white}
				 \vspace{1.7cm}
				{\hspace{1.1em}\fontsize{30}{60}\selectfont\textbf{Technical Report}} \\ [20pt]
				 \vspace{.2cm}{\hspace{1.1em}\huge\textbf{Paderborn University}} \\
				 \vspace{.2cm}{\hspace{1.1em}\huge\textbf{tr-ri-19-360}} \\
				 \vspace{.2cm}{\hspace{1.1em}\huge\textbf{\today}} 
				\vspace{1.5cm}
			}
		\end{minipage}%
		\begin{minipage}{.32\textwidth}
  				\includegraphics[width=4.8cm]{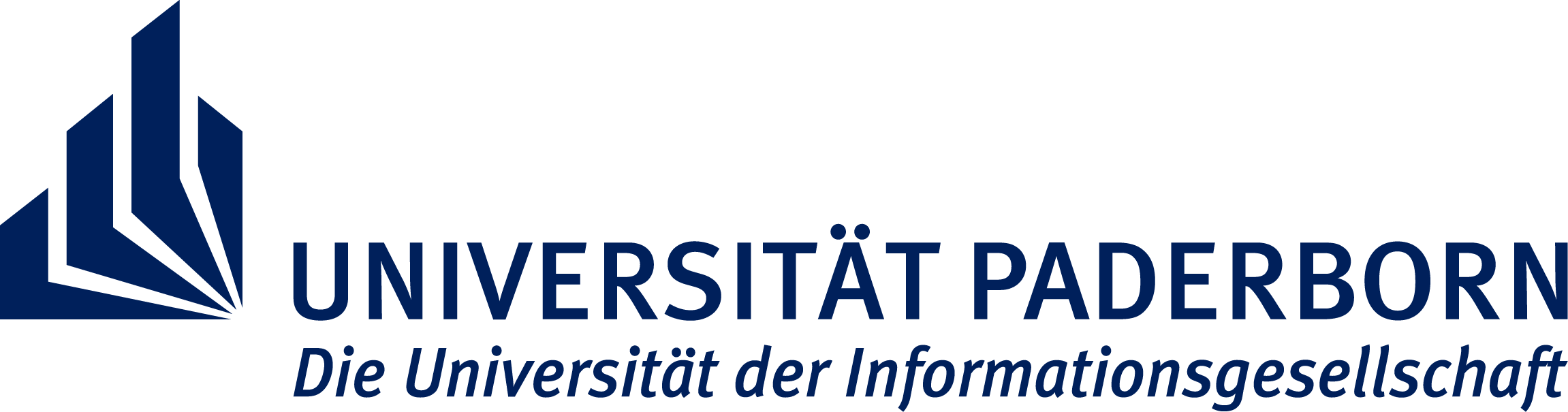} \\ [20pt]
  				\includegraphics[width=4.8cm]{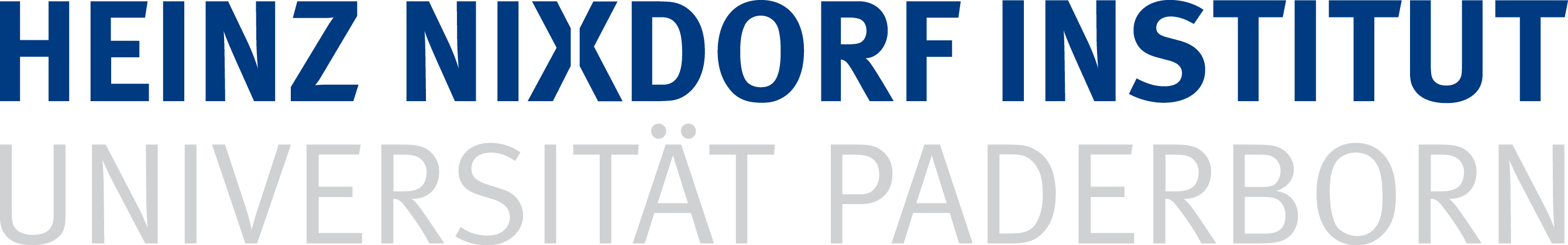}
		\end{minipage}%
		\end{minipage}
	}
		
	 \vspace{1.5cm}
	{\fontsize{20}{60}\selectfont\textbf{Object-Capability as a Means of Permission}} \\[10pt]
	{\fontsize{20}{60}\selectfont\textbf{and Authority in Software Systems}} 
	 
	\vspace{2cm}
	\includegraphics[height=4.5cm]{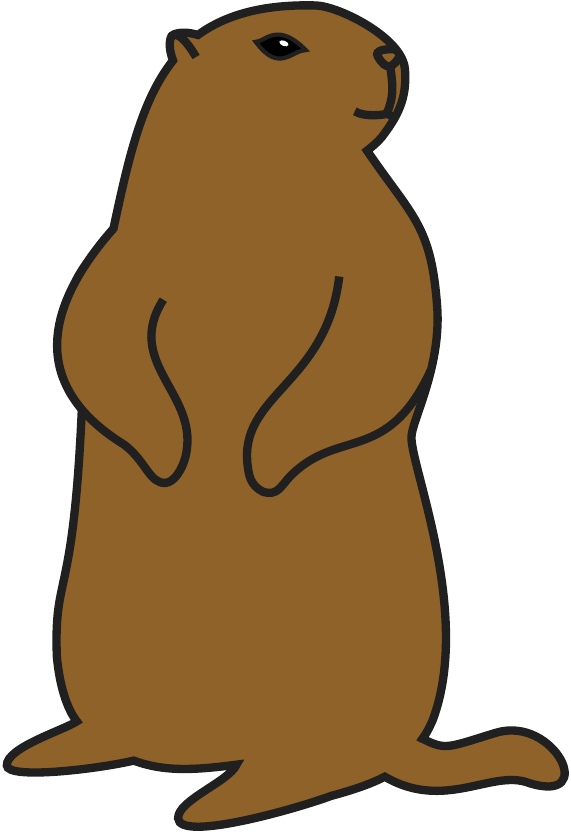}
	\vspace{2cm}
	 
	\begin{minipage}{17cm}
  		{\Large
			\textbf{Author:} \\ [5pt]
			{J\"{o}rn Koepe (Paderborn University)} \\[5pt]
		}
	\end{minipage}	
	
	\vspace{.2cm}
	\colorbox{ncs}{
		\begin{minipage}{17cm}
			{ \hspace{17cm}
			\vspace{.5cm}}
		\end{minipage}
	}

\end{center}
\newpage

%% file: text/abstract.tex
\begin{abstract}

The object-capability model is a security measure that consists in encoding access rights in individual objects to restrict its interactions with other objects. Since its introduction in 2013, different approaches to object-capability have been formalized and implemented. In this paper, we present the object-capability model, and present and discuss the state-of-the-art research in the area.
In the end, we conclude, that object capabilities can help in increasing the security of software, although this concept is not widely spread.

\end{abstract}

%% file: text/body.tex
\input{text/chapters/Introduction}
\input{text/chapters/Object-Capabilities}
\input{text/chapters/Comparision}
\input{text/chapters/Summary}

%% file: text/chapters/Introduction.tex
\section{Introduction}

Nowadays, many applications rely on external libraries to function correctly. However, including third-party libraries also opens an attack surface. Not all third-party programs are trustful or secure. A means to reduce security risks is to introduce access-control mechanisms, such as the object-capability model, introduced by Mark Miller in his doctoral dissertation~\cite{miller_robust_2006}.

In the object-capability model, which is designed for object-oriented languages, objects are used to manage access control. Specific objects can only access the behaviour they need, if they have a direct reference to the object providing that behaviour, following the terms of authority and permission. Those terms are used to model the access control, with \emph{permission} being the right to invoke certain behaviour on the targeted object, while \emph{authority} describes the ability to cause effects on an object~\cite{drossopoulou_permission_2016}.

In the past few years, new research was conducted on different areas of object-capabilities. Some focus on the formalism of the model~\cite{drossopoulou_permission_2016}, while others concentrate on analyzing security properties in that system~\cite{murray_analysing_2010}. Another common research topic is the study of typical problems of object-capabilities~\cite{drossopoulou_need_2013, albert_how_2014}.
Although capability safety was proven to imply authority safety~\cite{maffeis_object_2010}, the formal characterization of object capabilities
is not sufficient for verifying an application~\cite{devriese_reasoning_2016}. Nonetheless, it is possible to check for violations of object-capabilities with the help of model checking~\cite{rhodes_dynamic_2014}.

In this paper, we explain the concept of object-capability model, along with background information on related topics such as access control, authority, and permission. We then detail recent publications on object-capability, first looking at papers that focus on the formal aspect of object-capability, like introducing and analyzing different concepts of object-capability. Second, we present papers which focus on the implementation of object-capabilities and the advantages of the model for applications at a whole. Finally, the last category we explore includes papers that take a different approach to object-capability, for example, by analysing the corresponding model~\cite{albert_how_2014}  or how to automatically check for flaws in the model with the help of a model checker~\cite{rhodes_dynamic_2014}.
In the last part of the paper, we summarize the current state of the object-capability model, and discuss open questions on the topic.

%% file: text/chapters/Object-Capabilities.tex
\section{The Object-Capability Model}

In this section, we explain the object-capability model, and the notions of authority and permission.

\subsection{Object-Capabilities}

\begin{figure}[t]
        \centering
        \begin{tikzpicture}
        \node[state](s0) {A};
        \node[state, below of = s0, left of= s0](s1) {B};
        \node[state, below of = s0,right of = s0](s3){D};
        \node[state,  right of = s3, node distance = 3em] (s2) {C};

        \path[->]
        (s0) edge(s1)
        (s0) edge (s3);
        \end{tikzpicture}
        \caption{Object-capabilities from Listing~\ref{lst:objectCapability}. Edges represent references between the different objects of the program.}
        \label{fig:objectCapability}
    \end{figure}
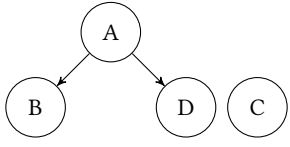

The \emph{object-capability model} was introduced by Mark Miller in his doctoral thesis in 2006~\cite{miller_robust_2006}. This security model is used to control accesses to particular parts of a program, in order to restrict potential malicious behaviour.
A \emph{capability} is defined as a token that describes access-rights, and \emph{object-capabilities} as capabilities applied to object-oriented languages. In the object-capability model, this token is thus a reference to a specific object.
An object can only interact with another object if it has the capability for the specific object. The example shown in Listing~\ref{lst:objectCapability} and Figure~\ref{fig:objectCapability} shows a simple implementation of the model.

\begin{figure}[t]
    \begin{lstlisting}[caption=Simple example for object-capabilities., label=lst:objectCapability]
main(){
    B B = new B()
    A A = new A(B) //  A -> B
    C C = new C()
    A.B.doSomething()
}

A {
    B;
    D:
    new A(B){
        B = B
        D = new D() // A -> D
    }
}
    \end{lstlisting}
\end{figure}

The object-oriented program in Listing~\ref{lst:objectCapability} contains three objects: \texttt{A}, \texttt{B}, and \texttt{C}. \texttt{A} receives a reference to \texttt{B} in its constructor, while \texttt{B} and \texttt{C} do not get any references. Following the object-capability model, \texttt{A} is allowed to call methods on \texttt{B}, while the other two objects can only use their own methods. In addition, \texttt{A} also has a reference to \texttt{D}, since it creates the object. References are encoded in a \emph{reference graph}, which also represents the given capabilities. The reference graph for Listing~\ref{lst:objectCapability} is shown in Figure~\ref{fig:objectCapability}.

Note that no matter at which point of the program execution the capability is created, the capability is valid over the entire execution, and the object owning a reference to another can invoke behaviour on that latter object.
There are multiple ways to gain capabilities for an object:
\begin{itemize}
\item The easiest one is to create a new instance of the desired object. By creating the object the creator object receives the reference and therefore the capability for the created object. In the example in Listing~\ref{lst:objectCapability} this happens for \texttt{A} and \texttt{D}, where the former gets a reference to the later by creating it.
\item It is also possible to introduce objects to one another. In the example, \texttt{A} could have introduced \texttt{B} to \texttt{C} or the other way around.
\item Another way to get a capability is to be born with it. This is also shown in the example, with \texttt{A} being constructed with a direct reference to \texttt{B}.
\end{itemize}

\subsubsection{Capability-safety}
A language is \emph{capability-safe} with respect to object-capability if it is restricted in a way that enforces the object-capability model. The most important restriction is that there should be no globally accessible mutable states~\cite{devriese_reasoning_2016}. This restriction enforces all communication between objects to be done over the capabilities.
Many common object-oriented languages, such as Java, can be used in a capability-safe way if these restrictions are met in the whole program. In the case of Java, this means not using static variables, since they are shared between all instances of a class.

\subsection{Authority and Permission}
The notions of \emph{authority} and \emph{permission} are broadly used to discuss the properties of the object-capability model. These terms were first introduced informally by Mark Miller along the object-capability model itself~\cite{miller_robust_2006}. He formalized them later together with Drossopoulou et al.~\cite{drossopoulou_permission_2016}.

To argue about permission and authority on a formal level, we first introduce notations for program states, following Drossopoulou et al.'s paper~\cite{drossopoulou_permission_2016}.

\subsubsection{Background}
In the following, we assume a capability-safe language, i.e., we assume no global variables and that the behaviour of an object can only be called if the caller has the reference to the specific object. We also assume a \emph{small step operational semantic}, meaning that only one computation step can be executed at a time, which we formalize as follows:
\[ P \vdash \sigma, stmts \rightsquigarrow \sigma', sr \]
where $\sigma$ is the state of the program at runtime, and $stmts$ are statements of the program. $sr$ is either a new statement or result.
For example, for the \texttt{Null} state, any number or object can be a result. An example language that fits the requirements as well as the definitions for the states, is detailed in Drossopoulou et al.'s paper~\cite{drossopoulou_permission_2016}.

$\sigma \rightsquigarrow^?\sigma'$ is called the \emph{liberal execution}, and describes every possible program execution.
In other words, it is the set of every possible computation sequence starting in $\sigma$ and resulting in $\sigma'$.

Moving on to the operational semantics, the behaviour of the program is modelled in a similar way.
We define the \emph{behavioural description} $B$, such that
\[ B \vdash \sigma, stmts \rightsquigarrow \sigma ', sr \]
where $B$ ensures that the liberal execution holds.
This description corresponds to a large range of behaviour such as low-level mechanisms or policy specifications~\cite{drossopoulou_permission_2016}.

Since all executions of a program are characterized by $B$, then $B$ also characterizes the program, which leads to the following definition:
\begin{definition}
    \begin{flalign*}
P\subset B &\Leftrightarrow \forall \sigma, \sigma', stmts, sr: P\vdash \sigma, stmts \rightsquigarrow \sigma', sr\\
&\implies B\vdash \sigma, stmts \rightsquigarrow \sigma ', sr
    \end{flalign*}
\end{definition}

The definition, the small step operational semantic, the liberal execution, and the behavioural description allow us to construct three so-called \emph{future worlds}, which describe how the program will look like in the next state.
These worlds, called the \ac{EW}, the \ac{BW}, and the \ac{MW}, are defined as follows:
\begin{definition}
\label{def:worlds}
    \begin{flalign*}
    EW(P,\sigma) &\equiv  \{\sigma' | \exists \ stms, sr: P \vdash \sigma, stmts\rightsquigarrow^*\sigma', sr\}\\
    BW(B,\sigma) &\equiv  \{\sigma' | \exists \ stms, sr: B \vdash \sigma, stmts\rightsquigarrow^*\sigma', sr\}\\
    MW(\sigma) &\equiv  \{\sigma' | \sigma \rightsquigarrow^{?*} \sigma'\}
    \end{flalign*}
\end{definition}
As stated in the Definition~\ref{def:worlds}, \acs{EW}s are reachable by a program, \acs{BW}s are created following a behavioural description, and \ac{MW}s are created by the liberal execution.
With these definitions, Drossopoulou et al. derive the following inclusions:
\begin{lemma}
    $\forall $ program $P$, behavioural description $B$ and state $\sigma$ :\begin{itemize}
        \item $\sigma \subseteq EW(P, \sigma) \subseteq MW(\sigma)$
        \item $BW(P, \sigma) \subseteq MW(\sigma)$
        \item $P \subset B \implies EW(P, \sigma) \subseteq BW(B, \sigma)$
    \end{itemize}
\end{lemma}
The full proof is found in Drossopoulou et al.'s paper~\cite{drossopoulou_permission_2016}.

\subsubsection{Permission}

The concept of \emph{permission} is described as the right to access an object directly and invoke its behaviour~\cite{miller_robust_2006, drossopoulou_permission_2016}.
With the definitions of the worlds, it is possible to create four different types of permission.

The first permission type is the base permission called \ac{CP}.
\begin{definition}
    \begin{gather*}
    CP(\sigma, o) \equiv \{o\} \cup \{ o' \exists f: \sigma(o,f) = o'\} \ \cup \\ \{o' | o = \sigma(this) \wedge \exists x.\sigma(x) = o'\}
    \end{gather*}
\end{definition}
Where $o$ is an object. The base permission states that if a subject has a direct access right to the object, it can invoke its behaviour. $\sigma(this)$ is a receiver of a currently executed method.

With the base permission, it is possible to define the three other types of permissions: the \ac{EP}, the \ac{BP}, and the \ac{MP}:
\begin{definition}
    \begin{flalign*}
    EP(P, \sigma, o) \equiv \bigcup_{\sigma' \in EW(P,\sigma)} CP(\sigma',o) \cap dom(\sigma)\\
    BP(B, \sigma, o) \equiv \bigcup_{\sigma' \in BW(P,\sigma)} CP(\sigma',o) \cap dom(\sigma)\\
    MP(\sigma, o) \equiv \bigcup_{\sigma' \in MW(P,\sigma)} CP(\sigma',o) \cap dom(\sigma)
    \end{flalign*}
\end{definition}
\ac{EP} are the permissions that are eventually created by a program $P$, \ac{BP} are those created by a behavioural description and \ac{MP} are those created by liberal executions.

Since the permissions are derived from the definition of worlds, we derive the following lemma:
\begin{lemma}
    $\forall$ program $P$, behavioural description $B$, and state $\sigma$:\begin{itemize}
        \item $CP(\sigma, o) \subseteq EP(P,\sigma, o)\subseteq MP(\sigma,o)$
        \item $CP(\sigma o) \subseteq BP(B, \sigma, o) \subseteq MP(\sigma,o)$
        \item $P \subset B \implies EP(P,\sigma, o) \subseteq BP(B,\sigma, o)$
    \end{itemize}
\end{lemma}

\subsubsection{Authority}
In contrast to permission, authority describes the possibility to invoke behaviour on an object
with out a direct link.
Similarly to permission, we can define four types of authority: the base \ac{CA}, the \ac{EA}, the \ac{BA}, and the \ac{MA}.

\begin{definition}
    \begin{align*}
    CA(P, \sigma, o) \equiv & \{ o'| \exists \sigma'', m, o_1, \ldots, o_n. \\
    & o \in CP(\sigma, \sigma(this)) \wedge\\
    & \forall i \in \{1..n\}.o_i \in CP(\sigma, o) \wedge \\
    & P \vdash \sigma'', x.m(x_1, \ldots x_n)\rightsquigarrow^*\sigma', \_ \wedge\\
    & \sigma'' = \sigma[x \mapsto, x_1\mapsto o_1,\ldots x_n \mapsto o_n] \wedge \\
    & \sigma(oÄ, f) \neq \sigma'(o', f) \}
    \end{align*}
    \begin{flalign*}
    EA(P, \sigma, o) \equiv \bigcup_{\sigma' \in EW(P,\sigma)} CA(\sigma', o) \cap dom(\sigma)\\
    BA(B,\sigma, o) \equiv \bigcup_{\sigma' \in BW(P,\sigma)} CA(\sigma', o) \cap dom(\sigma)\\
    MA(\sigma, o) \equiv \bigcup_{\sigma'\in MW(P,\sigma)} CA(\sigma', o) \cap dom(\sigma)
    \end{flalign*}
\end{definition}
With this definition, the \ac{CA}(o) permission contains every object which might be altered by calling a method. This means that the authority has the ability to modify an object. In this case, the subject does not need to have a direct permission to the object to be altered, as it can happen over other method calls.
Like for permissions, the following holds:
\begin{lemma}
    $\forall$ program $P$, behavioural description $B$, and state $\sigma$:\begin{itemize}
        \item $CA(\sigma, o) \subseteq EA(P,\sigma, o)\subseteq MA(\sigma,o)$
        \item $CA(\sigma o) \subseteq BA(B, \sigma, o) \subseteq MA(\sigma,o)$
        \item $P \subset B \implies EA(P,\sigma, o) \subseteq BA(B,\sigma, o)$
    \end{itemize}
\end{lemma}

\begin{figure}[t]
        \centering
        \begin{tikzpicture}[node distance = 5em]
        \node[state] (s0) {a};
        \node[state, left of=s0] (s1) {b};
        \node[state, below of=s1] (s2) {c};
        \node[state, left of=s1] (s3) {d};

        \path[->]
        (s1) edge (s0)
        (s2) edge (s0)
        (s3) edge (s1)
        (s3) edge [dashed, bend left] (s0)
        (s1) edge [dashed, bend right] (s0)
        ;

        \end{tikzpicture}
        \caption{Access path from Listing~\ref{lst:example}. Solid lines are indicate permissions, dashed lines indicate authority.}
        \label{fig:access}
    \end{figure}
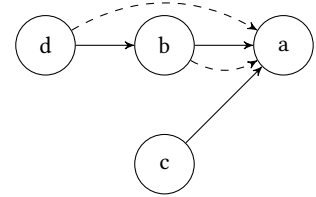

With these definitions, it is possible for an object to have authority but not permission over another object, and vice versa. We illustrate this in the example in Listing~\ref{lst:example} and Figure~\ref{fig:access}.
    \begin{figure}[t]
        \begin{lstlisting}[caption=Example program for authority and permission., label=lst:example]
class A(){
    int x;
}

class B(){
    A a
    B(A a) { this.a = a; }  // b --> a
    inc() { a.x ++; }
}

class C() extends B{
    inc(){};
}

class D(){
    B b;
    D(B b){ this.b = b; } // d --> b
    inc(){ b.inc(); }
}

main(){
    a = new A();
    b = new B(a); // b -> a
    c = new C(a); // c -> a
    d = new D(b); // d -> b
}
        \end{lstlisting}
    \end{figure}

As shown in the graph, \texttt{b} has permission and authority over \texttt{a}, since it has a direct reference to it and also can modify \texttt{a}. On the other hand, \texttt{c} can not modify \texttt{a}, as it does have authority over it. It has therefore no authority over \texttt{a}, although it has the permission on it. Finally, \texttt{d} can modify \texttt{a} without knowing it directly: by calling the \texttt{increase()} method from \texttt{b}, which in turn modifies \texttt{a}. So, \texttt{d} has authority over \texttt{a}, but no permission.

%% file: text/chapters/Comparision.tex
\section{Past Research on Object-Capability Approaches}

In this section, we detail past research on the topic of object-capability. We first present papers focusing on formalization, then on implementation, and finally, on the papers that do not fit well into the those sections.

\subsection{Formal Aspect of Object-Capability}
Many papers address the formalization of the object-capability model, with the goal of proving particular security properties with the help of object-capability.

In 2003, Miller et al.~\cite{miller_capability_2003} presented common myths on the topic of capabilities in their paper``Capability Myths Demolished''~\cite{miller_capability_2003}. In their paper, they address three myths: the \emph{equivalence myth}, the \emph{irrevocability myth} and the \emph{confinement myth}, which result form different interpretations of capability. The first myth states that access-control list systems and capability systems are equivalent, and the second, that capability-based access cannot be revoked. The last myth states that a capability system is not able to enforce confinement.
Miller et al. demonstrate that the three myths only hold for intermediate security models, but not for the pure capability model, or the object-capability model.

One of the most influential publications on the topic of object-capability is Mark Miller's doctoral thesis, called ''Robust Composition: Towards a Unified Approach to Access Control and Concurrency Control" ~\cite{miller_robust_2006}. The thesis introduces the object-capability model, and the core terms of authority and permission. Therefore this thesis can be considered as the foundation of object-capability. Miller developed these concepts on the basis of the already existing concept of capabilities and extended it to objects.

Another influential paper is ``Permission and Authority Revisited towards a formalisation''~\cite{drossopoulou_permission_2016} by Drossopoulou et al. In this paper, the concepts of authority, permission, and object-capability are extended and fully formalized, as this was left out in Miller's thesis.

Maffeis et al. show that capability safety implies authority safety in the paper``Object Capabilities and Isolation of Untrusted Web Applications''~\cite{maffeis_object_2010}. \emph{Authority safety} is defined with two properties: first, an object can only influence the authority of an object whose authority influences its own authority. Second, the change of authority for an object is bounded by the authority of the acting object. These properties are enough to ensure isolation and, as a result, can be used to isolate untrusted code.

In 2010, Murray promoted the construction of patterns to enforce security properties in his thesis ``Analysing Security Properties of Object-Capabilities''~\cite{murray_analysing_2010}. To describe a pattern, he uses Communicating Sequential Processes (CSP) and analyzes it with the help of a refinement-checker. In his thesis, Murray shows that it is possible to not only describe the properties of object-capabilities with CSP, but also the capabilities themselves. It is thus possible to reason about a CSP model for object-capability as a substitution for the real system, making it easier to prove the properties of the system.

Devriese, Birkedal and Pessens focus on the formal characterization of object-capabilities in their paper ``Reasoning about Object Capabilities with Logical Relations and Effect Parametricity''~\cite{devriese_reasoning_2016}. In this paper, they show that existing formal characterizations do not capture capability safety, and therefore it is not sufficient for verifying application security. To address this problem, they introduced a relation that helps to deliver a better characterization.

The paper ``Capabilities for Uniqueness and Borrowing'' by Haller and Odersky~\cite{haller_capabilities_2010} states that it is important for a system of unique objects to send messages over references, but the message is restricted due to the ill-effects of aliases. To solve the aliasing problem, they introduce a new type of uniqueness based on capabilities. This uniqueness enforces an at-most-one consumption of references and a notion of flexibility. To show that the new uniqueness is sound, the authors implemented and verified their approach through a simple example.

Swasey et al. introduce a logic called Object Capability Pattern Logic (OCPL), to describe object-capabilities in their paper ``Robust and Compositional Verification of Object Capability Pattern''~\cite{swasey_robust_2017}. Their logic, makes it easier to prove the robustness of programs.

The paper ``On Access Control, Capabilities, Their Equivalence and Confused Deputy Attacks'' by Rahani, Garg, and Rezk~\cite{rajani_access_2016} addresses the differences between access control, and capabilities and states, and argue that that the two notions are fundamentally different. Furthermore, they prove that it is not possible for capabilities to prevent all attacks, and therefore a capability-based system cannot be fully secure.

\subsection{Implementational Aspect of Object-Capability}

A large share of the related work also presents how to implement object-capability or how to ease their implementation.

Drossopoulou et al. present ``The Need for Capability Policies''~\cite{drossopoulou_need_2013} and ``How to Break the Bank: Semantics of Capability Policies''~\cite{albert_how_2014}. Both papers state that since the policies for object-capability are mostly only implicitly given, the code for the capabilities is often tangled with the main program code. The solution they propose is to make the policies explicit, with the programmer specifying their expectations about the security properties of the program.
In the first paper~\cite{drossopoulou_need_2013}, the authors state that this system would make it easier to reason about the explicit policies, and to check them for flaws.
In addition, the second paper~\cite{albert_how_2014} states that a structured specification is needed to express the policies and started designing such a specification~\cite{albert_how_2014}.

Another approach regarding the implementation of object-capability was presented by Darya Mellicher in her thesis proposal ``Controlling Module Authority via Programming Language Design''~\cite{melicher_controlling_2018}. This approach proposes the use of a capability safe module system for access-control.
In the proposal, Mellicher draws a first draft for such a system, which she and her co-authors also present in the paper ``A Capability-Based Module System for Authority Control"~\cite{melicher_capability-based_2017}. In this module system, each module is a statically typed capability.

\subsection{Other Focuses}

We now present the papers which use object-capability or propose similar approaches.

``Swapsies on the Internet''~\cite{drossopoulou_swapsies_2015} and ``Reasoning about Risk and Trust in an Open World'~\cite{drossopoulou_reasoning_2015}, both published by Drossopoulou et al. define the concept of risk and trust, and base their examples on top of the object-capability model . For those examples, the authors prove that the specifications for risk and trust are fulfilled.

An approach similar to object-capabilities is \emph{denied capabilities}, published in the paper ``Deny Capabilities for Safe, Fast Actors'' by Clebsch et al.~\cite{clebsch_deny_2015}. This model introduces a flexible type system, which can be used to deny control, in a similar system to access control.

The paper ``Minimal Ownership for Active Objects'' by Clarke et al~\cite{ramalingam_minimal_2008} states that it is natural for object-oriented languages to use active objects as an approach to concurrency. Active objects consist of an unshared state and a thread for control. The data sharing should only be done by references, but this leads to aliasing problems. To counter these problems, the authors prove that it is possible to pass objects belonging to one active object to another active object without copying the state. Although this paper does not deal with object-capabilities per se, the concept of data sharing over references is quite similar.

The last approach to object-capability is a model checker published in ``Dynamic Detection of Object Capability Violations Through Model Checking'' by Rhodes et al.~\cite{rhodes_dynamic_2014}. With the help of their model checker, the authors allow the visualization of leaks in a system.

%% file: text/chapters/Summary.tex
\section{Discussion and Conclusion}

Through the large body of research conducted on the topic, we see that object-capability has the potential to improve the security of distributed programs. However, even if the potential for security is there, it is not widely used, mostly due to convenience.
Global and static variables--which break the requirements for the object-capability model--are commonly used, especially for short scripts. Excluding their use from the beginning of a project could be an efficient solution.

Most publications on the topic of object-capability focus on the formal properties of the model, introducing extensions to the base model, analyzing its properties, or describing analysis methods. Current research agrees on the security of object-capability, although restrictions to the model are required, in some cases.

Other publications focus on implementation approaches to object-capabilities. The main two approaches are to make the policies for the model explicit (requiring a different program design to divide the code between object-capabilities and program behaviour), and to include object-capability modules directly in the programming language. This last approach is fairly new and only proposes early concepts for such a programming language.

Since the object-capability model is not included in most object-oriented languages, future work to encourage its adoption could be possible to provide a library of universal patterns. With the help of such patterns, software developers could implement the model in most object-oriented programming languages. This approach would require the developer to know about the model and the patterns, which could be achieved by including documentation in the style guides of the target programming languages.

All in all, object-capability has a high potential for enhancing software security. However knowledge about this technique is not widespread and object-capability is therefore rarely used. We encourage researchers and practitioners to spread the word about the potential of the object-capability model, and aim to include it into style guides and pattern lists. 